\documentclass[12pt,a4paper]{article}
\usepackage[biblatex]{anziamjedraft}

\newcommand{\pfrac}[2]{\frac{\partial #1}{\partial #2}}
\newcommand{\e}{\mathrm{e}}
\newcommand{\ii}{\mathrm{i}}
\DeclareMathOperator{\sgn}{sgn}

\title{Capillary levelling of thin liquid films of power-law rheology}
\author{Michael.~C. Dallaston}
\address{School of Mathematical Sciences, Queensland University of Technology,  Brisbane, QLD~4000, \textsc{Australia}.}
\mailto{michael.dallaston@qut.edu.au}
\myorcid{0000-0001-8993-6961}

\date{\today}

\begin{document}

\maketitle

\begin{abstract}
We find solutions that describe the levelling of a thin fluid film, comprising a non-Newtonian power-law fluid, that coats a substrate and evolves under the influence of surface tension.  We consider the evolution from both periodic and localised initial conditions as separate cases.  Particular (similarity) solutions in each of these two cases exhibit the generic property that the profiles are weakly singular (that is, higher-order derivatives do not exist), at points where the pressure gradient vanishes.  Numerical simulations of the thin film equation, with either periodic or localised initial condition, are shown to approach the appropriate particular solution.
\end{abstract}

\tableofcontents

\section{Introduction}
\label{sec:intro}

The importance of free-surface lubrication or coating flows to industry and science is well documented~\cite{Craster2009,Oron1997,Weinstein2004}.  Examples of applications include the construction of smooth (or deliberately patterned) surfaces, as well as predicting the evolution of lava flows and ice sheets, among others.  The mathematical modelling of such phenomena using the lubrication approximation is also ubiquitous~\cite{Craster2009,Oron1997}; this approximation allows the full governing partial differential equations and interfacial boundary conditions to be collapsed to a single equation for the film thickness, with only boundary conditions in the horizontal coordinates needed.

For small-scale industrial applications in which surface tension is a dominant effect, fluids are frequently better modelled using a non-Newtonian, rather than Newtonian, rheological model.  A very popular model is the power-law model:
\begin{equation}
\tau = \alpha|\dot\gamma|^{n-1}\dot\gamma, 
\label{eq:powerLawRheology}
\end{equation}
where $\tau$ is the stress tensor, $\dot\gamma$ is the strain rate tensor, $n$ is the power-law rheology exponent, and $\alpha$ is a constant of proportionality (see \citet{Myers2005} for a discussion of different rheological models).  The Newtonian case is recovered when $n=1$, in which case $\alpha$ is the fluid viscosity.  When $n<1$, the effective viscosity is decreasing in the strain rate, while for $n>1$, it is increasing.  These two cases are thus known as shear-thinning and shear-thickening, respectively.  Shear-thinning fluids are more common in practice.  The importance of rheology in the coating industry is described in \citet{Eley2005}.  

From a mathematical perspective, the evolution of a lubricating film, comprising a power-law fluid, under the effect of surface tension, has been considered with a focus on droplet spreading \cite{King2001,Rafai2004},  and on inclined or vertical substrates \cite{Allouche2015,Balmforth2003,Sylvester1973}.  The competition between surface tension and other effects such as van der Waals disjoining pressure, \cite{Garg2017}, and horizontally induced thermocapillary stress \cite{Mantripragada2022}, have also been considered with the use of numerical simulations.

In this article we focus on the deceptively simple problem that is the levelling of a film due to surface tension, in which a perturbed uniform film tends to flatten out over time.  In the Newtonian case, this phenomenon has been studied in the context of viscometry \cite{Benzaquen2013, Benzaquen2014, McGraw2011, McGraw2012}; by measuring the evolution of a perturbed film over time, the viscosity of the fluid may be determined.  In these works, explicit similarity solutions are found that model the levelling process.  This same application in the power-law fluid case has been studied to a lesser extent; \citet{Iyer1996} perform numerical calculations of the levelling problem with a Carreau fluid and compare to experimental results.  \citet{Ahmed2015} perform a purely numerical study on power-law fluids.  However, the explicit computation of similarity solutions that govern levelling behaviour, and comparison with numerical solutions, has not previously been undertaken for the power-law rheology.

Suitably nondimensionalised, the equation for a thin film of thickness $h(x,t)$ and rheology \eqref{eq:powerLawRheology} evolving purely due to surface tension in one dimension is~\cite{Garg2017,King2001}:
\begin{equation}
\pfrac{h}{t} + \pfrac{q}{x} = 0, \qquad q = h^{2+1/n}\left|\pfrac{^3h}{x^3}\right|^{1/n}\sgn\left(\pfrac{^3h}{x^3}\right),
\label{eq:thinFilmEq}
\end{equation}
where we have explicitly defined the flux $q$.  Here $n$ is the power-law exponent in the rheology; $n<1$ $n=1$, and $n>1$ correspond to the shear-thinning, Newtonian, and shear-thickening cases, respectively.  In the derivation of \eqref{eq:thinFilmEq}, the third derivative arises as the gradient of the Laplace pressure (the interface curvature being given by the second derivative in the lubrication approximation).
The governing equation \eqref{eq:thinFilmEq} has as an exact solution a perfectly uniform film which can be taken to be $h=1$ identically (we are free in the nondimensionalisation to specify the characteristic thickness as the vertical length scale).  

\begin{figure}
\centering
\includegraphics{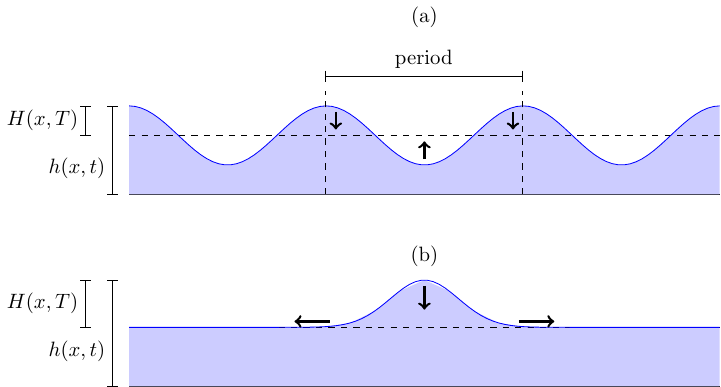}
\caption{A schematic diagram of the two levelling solutions under consideration: (a) levelling from an initially periodic perturbation of a flat film, described in Section \ref{sec:periodic}, and (b) levelling from an initially localised perturbation, described by the similarity solutions in Section \ref{sec:localised}.  In both cases $h$ describes the height of the film, while $H$ describes the difference from the uniform height of unity.}
\label{fig:schematic}
\end{figure}

In this work we consider the evolution of perturbations of a uniform film, distinguishing two cases: the evolution of a perturbation that is periodic in space, and the evolution of an initially localised perturbation in an infinite spatial domain (see Fig.~\ref{fig:schematic}).  The periodic case is equivalent to a perturbation over a finite domain with width equal to the period.
These two cases generalise known results for a Newtonian liquid film, namely, modal linear stability of a flat interface for periodic perturbations, and the similarity solution derived in \citet{Benzaquen2013, Benzaquen2014} for localised perturbations.  %
In the following section we will semi-analytically determine universal behaviour in each case as a perturbed film becomes level, before comparing with numerical simulations in Section \ref{sec:numerics}.

\section{Weakly singular nature and levelling solutions}
\label{sec:levelling}

\subsection{Weakly singular nature}
\label{sec:weaklySingular}
In the Newtonian case, the thin film equation \eqref{eq:thinFilmEq} is only singular at points where the thickness $h$ vanishes, which are not present in the study of levelling.  In the non-Newtonian case, \eqref{eq:thinFilmEq} is singular even if $h$ is not zero, at a point $x_0$ where the pressure gradient (that is, the third derivative of $h$) is zero.  We demonstrate this singular nature by expanding a solution near such a point.  A sensible expansion for $h$ near $x_0$ is
\[
h = h_0(t) + h_1(t)(x-x_0) + h_2(t)(x-x_0)^2 + h_r(t)|x-x_0|^r + \ldots, \qquad r \geq 3,
\]
where $r$ is an as-yet unknown power.  On substitution into \eqref{eq:thinFilmEq}, the leading-order term in the flux $q$ is 
\[
q = h_0^{2+1/n} \left[r(r-1)(r-2)|h_r| |x-x_0|^{r-3}\right]^{1/n}\sgn(h_r)\sgn(x-x_0) + \ldots
\]
so that
\[
\dot h_0 + h_0^{2+1/n} \left[r(r-1)(r-2) |h_r|\right]^{1/n} \left(\frac{r-3}{n}\right)\sgn(h_r) |x-x_0|^{(r-3)/n - 1} + \ldots = 0.
\]
Here $\dot h_0$ is the time derivative of the leading coefficient function.  Generally $\dot h_0 \neq 0$ at such a point, which requires $r = 3+n$.  For non-integer $n$, we expect the film profile will be weakly singular, in the sense that $h(x,t)$ is only $3+\lfloor n \rfloor$ times differentiable.  This is particularly important for shear-thinning flows ($n<1$) given the equation \eqref{eq:thinFilmEq} is fourth order.  We will observe this singular behaviour in each of the solutions constructed in this article.  We note that as special cases, if $\dot h_0 = 0$, then a singularity at higher order will occur as the gradient of flux must balance with another term, while if $n$ is an integer not equal to unity, then there will be higher-order noninteger powers in the flux that need to be balanced, requiring higher-order non-integer powers in the expansion of $h$.  We do not consider these special cases further here.

\subsection{Close-to-uniform approximation}

We examine the behaviour of a film that is close to the uniform solution $h=1$ by writing
\begin{equation}
h = 1 + \delta H(x, T), \qquad T = \delta^{1/n-1}t
\end{equation}
where $\delta \ll 1$.  Then to $O(\delta)$,
\begin{equation}
\pfrac{H}{T} + \pfrac{}{x}\left[\left|\pfrac{^3H}{x^3}\right|^{1/n}\sgn\left(\pfrac{^3H}{x^3}\right)\right] = 0.
\label{eq:thinFilmNearLevel}
\end{equation}
In the Newtonian case ($n=1$), \eqref{eq:thinFilmNearLevel} is linear, and no time rescaling is necessary, but in the non-Newtonian case ($n\neq 1$), \eqref{eq:thinFilmNearLevel} remains nonlinear, and the amplitude-dependent scaling in time is necessary for the two terms in the equation to be in balance.  The near-level approximation \eqref{eq:thinFilmNearLevel} has the same property of the original equation \eqref{eq:thinFilmEq}, in that $H$ must be weakly singular at any point where the third spatial derivative is zero in order for the evolution to be well defined.

We now construct two special classes of solutions to \eqref{eq:thinFilmNearLevel}, generalising the known results for the Newtonian case (see Fig.~\ref{fig:schematic}).  Firstly, we consider solutions periodic in $x$, which are relevant for initial conditions that are periodic or on bounded domains with no-flux conditions.  Secondly, we construct similarity solutions that are appropriate for initial conditions that are spatially localised.

\subsection{Periodic initial condition}
\label{sec:periodic}

We start with solutions that are periodic in $x$.  In the Newtonian case ($n=1$) this is equivalent to computing the linear stability of the uniform solution.  For a given wavenumber $k$ (so that the period is $2\pi/k$) we have
\[
H(x,t) = \bar A\e^{-k^4 t} \cos(kx),
\]
where the amplitude $\bar A$ is arbitrary (equivalent to time-translational invariance).  For $n\neq 1$, the near-level behaviour is essentially nonlinear, and linear stability approaches are not appropriate.  Instead, we assume a solution of \eqref{eq:thinFilmNearLevel} more generally in which the time and space dependence may be separated, or, equivalently, a similarity solution in which the horizontal spatial scale remains fixed.  The appropriate ansatz depends on whether the fluid is shear-thinning or thickening:
\begin{equation}
H(x,T) = \begin{cases}
\displaystyle \left(\frac{n}{1-n}\right)^{n/(1-n)} T^{-n/(1-n)} \bar H(x), & n < 1 \\
\displaystyle \left(\frac{n}{n-1}\right)^{-n/(n-1)} (-T)^{n/(n-1)} \bar H(x), & n > 1.
\end{cases}
\label{eq:periodicAnsatz}
\end{equation}
where $\bar H(x)$ is a periodic, but as yet undetermined, function (the $n$-dependent constant is included to simplify the problem for $\bar H$).  The above ansatz represents infinite-time but algebraic decay as $T\to\infty$ for the shear-thinning ($n < 1$) case, and finite-time (at $T\to 0^-$) levelling for the shear-thickening $n > 1$ case.  As written \eqref{eq:periodicAnsatz} is valid for $T>0$ and $T<0$ for $n<1$ and $n>1$, respectively; however, the solutions are of course invariant to translations in time, and in particular the finite levelling time will depend on the initial amplitude.

Under the ansatz \eqref{eq:periodicAnsatz}, $\bar H$ satisfies the ordinary differential equation
\begin{equation}
[|\bar H'''|^{1/n}\sgn(\bar H''')]' = \bar H
\label{eq:periodicODEstep}
\end{equation}
(where $'$ represents differentiation with respect to $x$), for both shear-thinning and thickening.  Let $U = |\bar H'''|^{1/n}\sgn(\bar H''')$, then on differentiating \eqref{eq:periodicODEstep} three times, $U$ satisfies
\begin{equation}
U'''' = |U|^n\sgn U,
\label{eq:periodicODE}
\end{equation}
with $\bar H$ readily found from $U$ by $\bar H = U'$.  Since any solution to \eqref{eq:periodicODE} may be scaled onto another solution by $x \mapsto \lambda x$, $U \mapsto \lambda^{4/(1-n)}U$, it suffices to find a solution with period $2\pi$, which can then be scaled to find the solution for any other period.

\subsubsection{Numerical computation}

Solutions to \eqref{eq:periodicODE} will be weakly singular at points $x_0$ where $U(x_0)=0$, with an expansion of the form
\[
U = C_1(x-x_0) + C_2(x-x_0)^2 + C_3(x-x_0)^3 + C_r|x-x_0|^{4+n}\sgn(x-x_0) + \ldots,
\]
where $C_1, C_2, \ldots$, are the coefficients in the series.  This singularity corresponds to the behaviour of the original problem near a singular point described in Section \ref{sec:weaklySingular} for noninteger $n$.  In particular, in the shear-thinning case, $U$ will be four (but not five)-times differentiable, corresponding to the profile $\bar H$ (and so the time-dependent profile $h$) being three but not four-times differentiable at points where the pressure gradient $h_{xxx} = 0$.  However, since \eqref{eq:periodicODE} is fourth order, this singularity is weak enough to be able to proceed (that is, we have essentially integrated the equation by solving for the antiderivative of $\bar H$).

Assuming a solution profile that is symmetric about a point $x=0$, a $2\pi$-periodic solution is equivalent to  finding a solution on the half period $0 < x < \pi$ that satisfies the conditions
\begin{equation}
U(0) = U''(0) = U(\pi) = U''(\pi) = 0.
\label{eq:periodicSymmetryBCs}
\end{equation}
In the half-period, $U$ may be taken to be positive.  
We calculate solutions to the boundary value problem \eqref{eq:periodicODE}, \eqref{eq:periodicSymmetryBCs} numerically using the MATLAB function \texttt{bvp4c}.  An initial guess is provided close to $n=1$ from the asymptotic solution described below (Section \ref{sec:periodicAsymptotics}).  A continuous branch of solutions is then constructed by numerical continuation for decreasing and increasing power-law exponent $n$, using the previously computed solution as an initial guess for the next.

\begin{figure}
\centering
\includegraphics{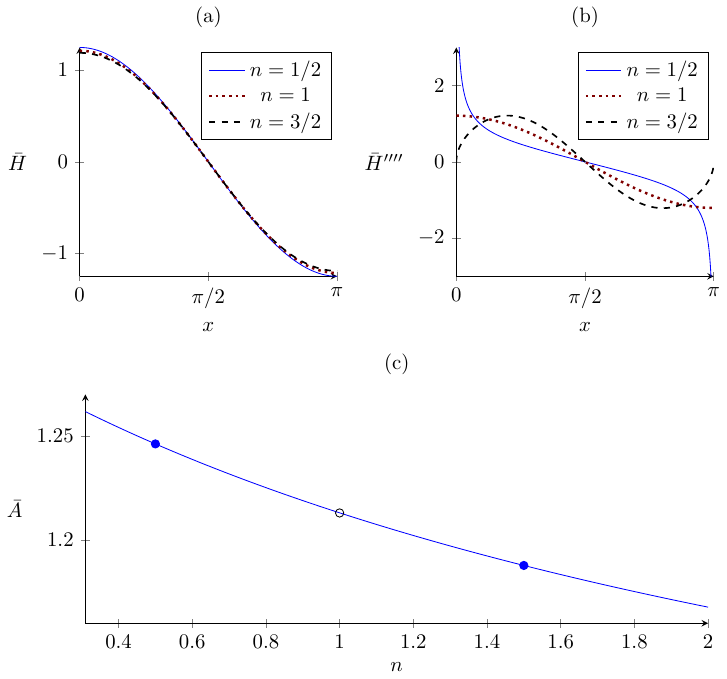}
\caption{Spatial profiles $\bar H(x)$ representing levelling solutions for a periodic initial condition, $n = 1/2$ (shear-thinning) and $n=3/2$ (shear-thickening), with the exact cosine profile of the Newtonian case $n=1$ shown for reference.  (a) The solution profiles (calculated from solutions $U$ to \eqref{eq:periodicODE} with $\bar H = U'$).  These profiles are close to sinusoidal.  (b) The fourth derivative $\bar H''''$ show the singular nature of these profiles, with $\bar H$ only three times differentiable for $n=1/2$, and only four times differentiable for $n=3/2$.  (c) The dependence of amplitude $\bar A$ on power-law exponent $n$.}
\label{fig:periodic}
\end{figure}

In Fig.~\ref{fig:periodic} we plot the results of this computation, focussing on the values $n=1/2$ and $n=3/2$.  These exponents are chosen as the examples of shear-thinning and thickening cases throughout our study.  We include both the profiles $\bar H = U'$, as well as the fourth derivative $\bar H'''' = n|U|^{n-1}U'$.  The profiles themselves are visually difficult to distinguish from the sinusoidal profiles that are exact for the Newtonian case ($n=1$), but have non-arbitrary, $n$-dependent amplitude.  In plotting the fourth derivative, the difference between the cases is stark, in particular, that the shear-thinning case ($n=1/2$) is singular at the peak and trough of the periodic profile ($x=0, \pi$), while for the shear-thickening case ($n=3/2$), the fourth derivative is bounded and continuous (although the next derivative will indeed be singular).  We also plot the amplitude $\bar A$, calculated as half of the peak to trough distance $\bar A = (\bar H(0) - \bar H(\pi))/2$; this value varies only weakly in the range of $n$ values calculated.

\subsubsection{Close-to-Newtonian limit}
\label{sec:periodicAsymptotics}
In the above method, the starting point of the numerical solutions was found by determining the asymptotic solution to \eqref{eq:periodicODE} in the close-to-Newtonian limit $n \to 1$.  This approximation is also of interest in itself.  Defining $n = 1+\epsilon$, where $\epsilon \ll 1$:
\[
U^n = U\e^{\epsilon\log U} = U(1 + \epsilon \log U + \tfrac{1}{2}\epsilon^2\log(U)^2 + \ldots),
\]
so that if $U = U_0 + \epsilon U_1 + \epsilon^2 U_2 + \ldots$, we have
\[
U_0'''' = U_0, \qquad U_1'''' - U_1 = U_0\log(U_0).
\]
Applying the boundary conditions \eqref{eq:periodicSymmetryBCs} to $U_0$, we find
\[
U_0 = \bar A\sin(x).
\]
The amplitude $\bar A$ is determined from a solvability condition for $U_1$.  Applying the same boundary conditions to $U_1$, we must have that (by the Fredholm alternative)
\[
\int_0^\pi U_0\log(U_0) \sin(x)\,\mathrm dx = \frac{\pi \bar A}{2}\left[\log \bar A + \frac{1}{2}(1-\log 4)\right] = 0.
\]
Thus either $\bar A=0$ (the trivial solution), or 
\begin{equation}
\bar A = \e^{(\log 4 - 1)/2} = 2\e^{-1/2} = 1.21\ldots
\label{eq:solvabilityA}
\end{equation}
This value agrees with the numerically computed results near $n=1$ (see Fig.~\ref{fig:periodic}).  We note that in this asymptotic result we have not considered the region near $U=0$ where $U$ will be weakly singular, as it does not play a role in selecting the amplitude $\bar A$.

\subsection{Localised initial condition}
\label{sec:localised}

We now find solutions relevant for an initial condition that is localised in an infinite spatial domain.  In this case, relevant solutions are similarity solutions that describe how an initial peak spreads over time.

For any value of $n$, similarity solutions to \eqref{eq:thinFilmNearLevel}, symmetric around the point $x=0$, take the form
\begin{equation}
H(x,t) = H_0T^{-\alpha}F(\eta), \qquad \eta = H_0^{(1-n)/(n+3)} \alpha^{-n/(n+3)}\frac{x}{T^\alpha}, \qquad \alpha = \frac{n}{4},
\label{eq:similarityAnsatz}
\end{equation}
where $\eta, F$ are the similarity variables, $H_0$ is an arbitrary amplitude included so that we can specify $F(0) = 1$, and the similarity exponent $\alpha$ is determined from the two conditions that the terms in \eqref{eq:thinFilmNearLevel} have the same time dependence, and that mass is conserved.  Substituting \eqref{eq:similarityAnsatz} into \eqref{eq:thinFilmNearLevel} results in the ordinary differential equation for similarity profiles:
\[
(\eta F)' = [|F'''|^{1/n}\sgn(F''')]',
\]
(here $'$ refers to derivatives with respect to $\eta$) subject to symmetry conditions at zero.  This equation may be integrated once and the boundary conditions used to result in the third-order equation
\begin{equation}
F''' = |\eta F|^n\sgn(F), \ \eta > 0, \qquad F(0) = 1, \ F'(0) = 0.
\label{eq:similarityODE}
\end{equation}
A third condition is imposed by requiring solutions to decay as $\eta\to\infty$.  We note that \eqref{eq:similarityODE} is a special case of a class of nonlinear differential equations whose existence is established in \citet{Bernis1991}, although they do not explicitly construct solutions numerically.  Similarly to the periodic case, since we only deal with third derivatives of $F$ in \eqref{eq:similarityODE}, the singularities that are present (in particular in the fourth derivative for $n<1$) will not prevent numerical computation of the solution.

\begin{figure}
\centering
\includegraphics{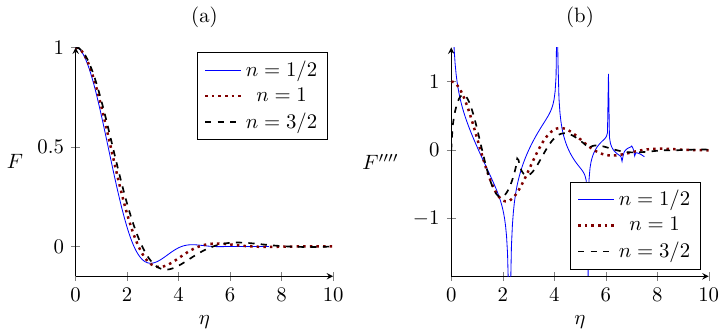}
\caption{Similarity solutions for thin film levelling of an initially localised perturbation, for power-law exponents $n=1/2$ and $n=3/2$, as well as the Newtonian case $n=1$.  While the solution profiles (a) appear qualitatively similar, plotting the fourth derivative (b) demonstrates singularities in the higher derivatives of the non-Newtonian similarity solutions.}
\label{fig:similarity}
\end{figure}

In the Newtonian case, the similarity solution has been previously identified in \citet{Benzaquen2013, Benzaquen2014}; the similarity exponent is $\alpha = 1/4$, and the equation for the profile \eqref{eq:similarityODE} is linear:
\begin{equation}
F''' = \eta F.
\label{eq:similarityODENewtonian}
\end{equation}
Given $F(0) = 1$, $F'(0)=0$, and decay at infinity, \eqref{eq:similarityODENewtonian} has an exact solution that may be expressed in terms of the Fourier integral
\begin{equation}
F(\eta) = C\int_{-\infty}^\infty \exp\left(\ii \xi \eta - \frac{\xi^4}{4}\right) \,\mathrm d\xi = C\eta^{1/3}\int_{-\infty}^\infty \exp\left[\left(\ii \zeta - \frac{\zeta^4}{4}\right)\eta^{4/3}\right]\,\mathrm d\zeta,
\label{eq:similarityFourierSolution}
\end{equation}
where $C = \sqrt{2}/\Gamma(1/4)$ is determined from the condition $F(0) = 1$.  The large-$\eta$ asymptotic behaviour of \eqref{eq:similarityFourierSolution} is readily determined from a steepest descent calculation, with a contour that passes through two critical points $\zeta = \e^{\ii\pi/6}, \e^{5\ii\pi/6}$, which ultimately results in
\[  
F(\eta) \sim \frac{4\sqrt{\pi}}{\sqrt{3}\Gamma(1/4)} \eta^{-1/3}\exp\left(-\frac{3}{8}\eta^{4/3}\right)\cos\left(\frac{3\sqrt 3}{8}\eta^{4/3} - \frac{\pi}{6}\right), \qquad \eta \to \infty.
\]
This asymptotic result indicates that $F$ decays in oscillatory fashion as $\eta\to\infty$.  In addition, the solution \eqref{eq:similarityFourierSolution} may be represented exactly in terms of hypergeometric functions:
\begin{equation}
F(\eta) = {}_0F_2\left([], [1/2, 3/4], \eta^4/64\right) - \frac{\Gamma(3/4)}{4\Gamma(5/4)} \eta^2 {}_0F_2\left([], [5/4, 3/2], \eta^4/64\right).
\label{eq:similarityNewtonianExactSolution}
\end{equation}
This solution is depicted in Figure \ref{fig:similarity}a.

Returning to the non-Newtonian case $n \neq 1$, numerical solutions to \eqref{eq:similarityODE} are obtained by shooting from $\eta=0$.  Since $F(0) = 1, F'(0) = 0$, there is one parameter $F''(0)$ that must be determined numerically by requiring that $F\to 0$ as $\eta \to \infty$.  We use \texttt{ode45} in MATLAB to solve the equation up to a sufficiently large value of $\eta$, combined with a root finding method to determine the appropriate value of $F''(0)$.  We perform this calculation for the exponents $n = 1/2$ and $n=3/2$ (for $n=1$, the numerical method matches the exact solution \eqref{eq:similarityNewtonianExactSolution}).  The solutions so found are plotted in Fig.~\ref{fig:similarity}, including both the profiles $F$ and the fourth derivative, calculated from the numerical solution using
\[
F'''' = n|\eta F|^{n-1}(F + \eta F').
\]
As was the case for periodic solutions, plotting this fourth derivative shows the presence of singularities at higher order.  Similar to the Newtonian case, the non-Newtonian similarity solutions oscillate an infinite number of times as $\eta\to\infty$, so there is an infinite sequence of such singularities.

\section{Comparison with numerical simulations}
\label{sec:numerics}

In order to test the universality of the solutions found in Section \ref{sec:levelling}, we simulate the time-dependent lubrication equation \eqref{eq:thinFilmEq}.
We use a finite volume method \cite{Patankar1980}, in which fluxes are explicitly computed.  This property is particularly important in the present case as for $n<1$, we do not expect \eqref{eq:thinFilmEq} to have four-times differentiable solutions, whereas the flux $q$ should be differentiable.  We consider a domain $x \in [0,L]$ divided into $N$ cells of width $\Delta x = L/N$, with the $j$th cell centre at $x_j = \Delta x (j-1/2)$, $1 < j < N$.  The finite difference expression for the third derivative of $h$ on the face between the $j$th and $(j+1)$th cell is
\[
\pfrac{^3h}{x^3} \approx \frac{1}{\Delta x ^3}\left(-h_{j-1} + 3h_{j} - 3h_{j+1} + h_{j+2}\right),
\]
where $h_j$ is taken to be the representative value of $h$ in cell $j$, while the value of the thickness $h$ itself on the face is given by the average $(h_{j+1} + h_j)/2$.  These expressions are used to approximate the fluxes $q_j$.  The transport equation \eqref{eq:thinFilmEq} then gives
\[
\frac{\mathrm d h_j}{\mathrm dt} = \frac{1}{\Delta x}\left(q_{j} - q_{j-1}\right).
\]
Zero-flux conditions are imposed at the two ends $x=0$ and $x=L$, while zero-slope conditions ($\partial h/\partial x=0$) are used to define ghost node values that allow the third derivative to be computed on all internal faces.  The numerical scheme is then advanced in time using MATLAB's \texttt{ode15s} implicit time-stepping algorithm.

\begin{figure}
\centering
\includegraphics{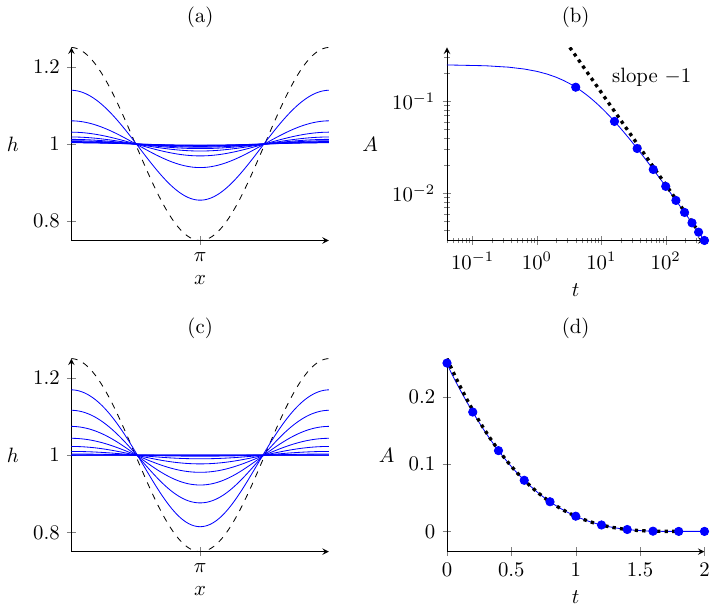}
\caption{Numerical simulation of a periodic solution for (a,b) $n=1/2$, and (c,d) $n=3/2$, using the method described in Section \ref{sec:numerics}.  (a,c) depict the profile evolution for each case, with initial condition shown as the black dashed curve.  In (b) we observe the predicted power-law decay in amplitude for shear-thinning flow, with $A \sim  1.246 t^{-1}$ (shown as a black dotted line).  In (d) we observe the predicted finite time levelling in the shear-thickening case, with $A \sim 0.044(t_0-t)^3$ (shown as a black dotted line); here the finite levelling time is approximately $t_0 \approx 1.8$.  Blue circles in (b,d) correspond to the times at which profiles are plotted in (a,c), respectively.}
\label{fig:periodicComparison}
\end{figure}

Using this method we compute solutions of \eqref{eq:thinFilmEq} for the two values of the power-law exponent we have focussed on thus far ($n=1/2$ and $n=3/2$), and initial conditions that test the behaviour of periodic and localised solutions, respectively.  To test the behaviour of periodic solutions, we choose a domain size of $L=2\pi$ and an initial condition
\[
h(x,0) = 1 + 0.25\cos\left(x\right).
\]
From the ansatz \eqref{eq:periodicAnsatz} we predict that the amplitude $A(t)$, which we calculate numerically as $(\max(h) - \min(h))/2$, will either decay algebraically or undergo finite-time levelling, depending on $n$:
\begin{equation*}
A \sim \begin{cases}  \displaystyle \bar A\left(\frac{n}{1-n}\right)^{n/(1-n)}t^{-n/(1-n)}, & n < 1 \\ \displaystyle \bar A\left(\frac{n}{n-1}\right)^{-n/(n-1)} (t_0-t)^{n/(n-1)}, & n > 1 \end{cases},
\end{equation*}
where $\bar A$ is the ($n$-dependent) amplitude of the solution to the profile function $\bar H(x)$, and $t_0$ is the finite levelling time.  For $n=1/2$ and $n=3/2$, then, these become
\begin{equation}
A \sim \begin{cases} 1.213 t^{-1}, & n = 1/2 \\ 0.044(t_0-t)^{3}, & n = 3/2 \end{cases}.
\label{eq:periodicAmplitudeCases}
\end{equation}
In this case, the prefactor is determined independently of the initial condition (although not of the domain size).  The levelling time $t_0$ is initial condition-dependent, however.  In Fig.~\ref{fig:periodicComparison} we show the results of the numerical simulation.  The observed behaviour of the amplitudes are in agreement with the prediction \eqref{eq:periodicAmplitudeCases}, indicating that the levelling solutions we found in Section \ref{sec:levelling} are accurate, and are attractors of more generic initial conditions.

\begin{figure}
\centering
\includegraphics{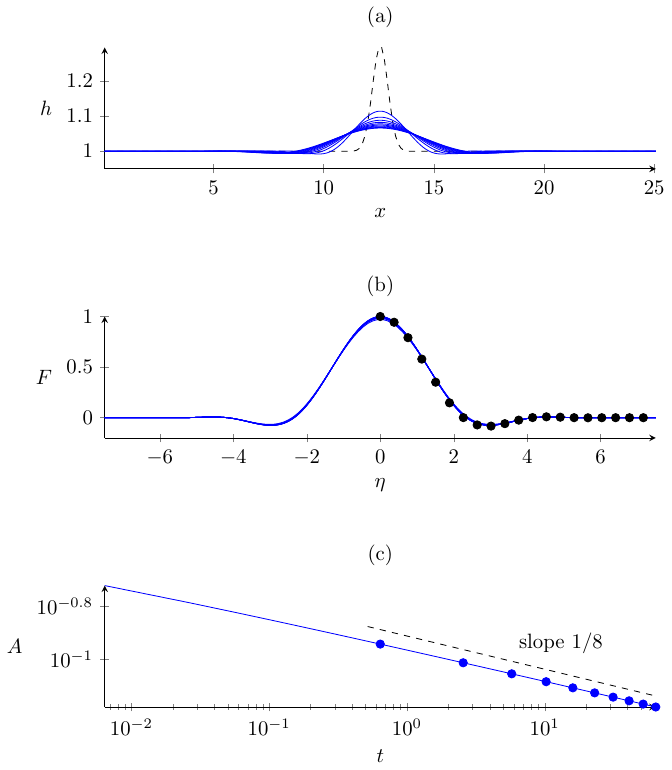}
\caption{Numerical simulation of a localised initial perturbation for the shear-thinning case ($n=1/2$).  (a) The solution profiles themselves; (b) the rescaled profiles (according to \eqref{eq:similarityAnsatz}) collapse onto a single curve which matches the similarity profile computed in Section \ref{sec:localised} (black circles).  (c) The power-law decay of the amplitude $A = \max(h) - 1$ also matches the value predicted by the similarity ansatz.  Blue circles in (c) correspond to the profiles plotted in (a,b).}
\label{fig:thinSimilarity}
\end{figure}

\begin{figure}
\centering
\includegraphics{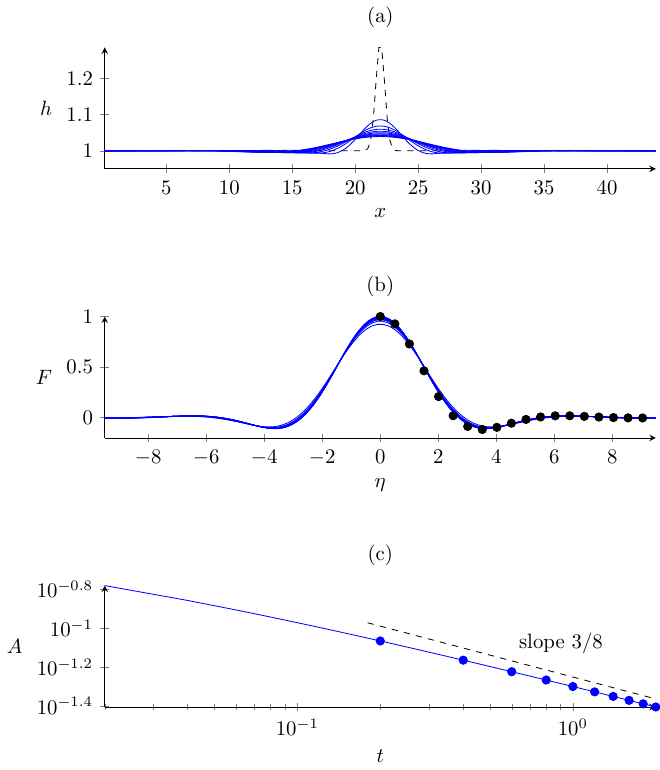}
\caption{Numerical simulation of a localised initial perturbation for the shear-thickening case ($n=3/2$), with, similar to Fig.~\ref{fig:thinSimilarity}: (a) The profiles, (b) the rescaled profiles which match with the similarity solution (black circles), and (c) the power-law decay of the amplitude $A$, which matches the predicted value.  Blue circles in (c) correspond to the profiles plotted in (a,b).}
\label{fig:thickSimilarity}
\end{figure}

To test the behaviour of localised initial conditions, we start with a Gaussian initial condition
\[
h(x,0) = 1 + 0.3\exp[-4(x-L/2)^2],
\]
for both $n=1/2$ and $n=3/2$, with domain size $L$ sufficiently large so as to avoid the boundary having an effect on the evolution.  The similarity ansatz \eqref{eq:similarityAnsatz} predicts the amplitude, defined by $A = \max(h) - 1$, will go as
\begin{equation}
A \sim \text{constant}\cdot t^{-n/4}
\label{eq:similarityAmplitude}.
\end{equation}
As $H_0$ is arbitrary in \eqref{eq:similarityAnsatz}, the prefactor in this relation is dependent on initial condition.  We plot the results of these simulations in Fig.~\ref{fig:thinSimilarity} for $n=1/2$, and Fig.~\ref{fig:thickSimilarity} for $n=3/2$.  In each case, the profiles rapidly evolve toward the relevant similarity solution (with the appropriate value of $H_0$ estimated from the numerical solution at the final time), and the behaviour of the amplitude $A$ approaches the predicted power law \eqref{eq:similarityAmplitude}.  Again, the strong agreement indicates the similarity solution is an attractor for initially localised perturbations on an infinite domain.

\section{Discussion}

By explicitly calculating similarity solutions that describe levelling, we have determined the power-law decay of the amplitude of perturbations of a flat film.  As the decay depends on the power in the rheology \eqref{eq:powerLawRheology}, our results imply that levelling experiments could be used to determine the rheology of fluids, in a similar way as they are used to determine the viscosity of presumed-Newtonian fluids \cite{Benzaquen2013,Benzaquen2014}.  

We have also demonstrated that solutions to lubrication equations with power-law rheology will be weakly singular at points where the pressure gradient vanishes.  This will be also true in more complicated models, for instance those that feature disjoining pressure, Marangoni forces, or gravity.   Despite the weakly singular nature of solutions to \eqref{eq:thinFilmEq}, previous studies have successfully computed solutions numerically.  Indeed, the distinct infinite-time and finite-time levelling, for shear-thinning and thickening fluids, respectively, can be observed in the numerical results of \citet{Ahmed2015}, although they do not interpret them as such.  Numerical studies on power-law thin films are likely to numerically regularise the singularities, or be formulated in such a way that they correctly solve the weak version of the equation (as our method does).  However, as the singular behaviour becomes more severe as $n$ is reduced, it is likely that it will be important to take into account for studies that wish to examine the highly shear-thinning limit ($n\to 0$).  On the other hand, studies on inclined or vertical planes, or with imposed tangent stress, may avoid this issue by never having points at which the pressure gradient vanishes.  

The rheology itself may be explicitly regularised, using, for example, a Carreau-type model \cite{Myers2005}, which acts like a power-law fluid except for small strain rate, where it behaves in a Newtonian manner.  For more complex rheology, the flux becomes more complicated to compute, although it is possible to make progress using the method described in \citet{Pritchard2015} (see also \citet{Hinton2022} for the two-dimensional case).  Under such a regularisation, a thin film may be expected to level in the same way as described in this paper, until the film becomes very close to level, at which point the Newtonian regime will be reached, and perturbations will switch to exponential decay according to the linear stability analysis of the Newtonian case.

Extending to two spatial dimensions is also of interest.  Presumably, there exists a radially symmetric similarity solution that would describe the decay of localised perturbations.  For periodic perturbations, one could imagine a radial solution for a thin film in a finite circular domain, or, for a more general shape, one would have to solve the nonlinear elliptic equation that is the counterpart to \eqref{eq:periodicODE} in higher dimensions.  The power-law decay of the amplitude in time, however, would be universal.

\printbibliography

\end{document}